\documentclass[twocolumn]{jpsj2}
\usepackage{graphicx}
\usepackage{amsmath,amssymb}

\title{ 
Superconducting Fluctuation and Pseudogap in 
\\ Disordered Short Coherence Length Superconductor 
} 

\author{Youichi {\sc Yanase}\footnote{E-mail:
yanase@itp.phys.ethz.ch}}

\inst{Department of Physics, University of Tokyo, Tokyo 113-0033, Japan \\
Theoretische Physik, ETH-Honggerberg, 8093 Zurich, Switzerland}

\recdate{August 17, 2006}

\abst
{ 
 We investigate the role of disorder on the superconducting (SC) 
fluctuation in short coherence length $d$-wave superconductors. 
 The particular intetest is focused on the disorder-induced 
microscopic inhomogeneity of SC fluctuation and its effect on the 
pseudogap phenomena. 
 We formulate the self-consistent $1$-loop order theory for the 
SC fluctuation in inhomogeneous systems and 
analyze the disordered $t$-$t'$-$V$ model. 
 The SC correlation function, electronic DOS and 
the critical temperature are estimated. 
 The SC fluctuation is localized like a nanoscale granular structure 
when the coherence length is short, namely the transition temperature 
is high. 
 This is contrasted to the long coherence length 
superconductors where the order parameter 
is almost uniform in the microscopic scale. 
 In the former case, the SC fluctuation is enhanced by the disorder 
in contrast to the Abrikosov-Gorkov theory. 
 These results are consistent with the STM, NMR and transport 
measurements in high-$T_{\rm c}$ cuprates and illuminate the essential 
role of the microscopic inhomogeneity. 
 We calculate the spacial dependence of DOS around the single impurity 
and discuss the consistency with the NMR measurements. 
}

\kword
{
Microscopic inhomogeneity; pseudogap; superconducting fluctuation; 
high-T$_{\rm c}$ cuprates; randomness
}

\begin{document}
\sloppy
\maketitle

\newcommand{\eli}{$\acute{{\rm E}}$liashberg }
\renewcommand{\k}{\mbox{\boldmath$k$}}
\newcommand{\q}{\mbox{\boldmath$q$}}
\newcommand{\Q}{\mbox{\boldmath$Q$}}
\newcommand{\kk}{\mbox{\boldmath$k'$}}
\newcommand{\e}{\varepsilon}
\newcommand{\ee}{\varepsilon^{'}}
\newcommand{\s}{{\mit{\it \Sigma}}}
\newcommand{\J}{\mbox{\boldmath$J$}}
\newcommand{\vv}{\mbox{\boldmath$v$}}
\newcommand{\Jh}{J_{{\rm H}}}
\newcommand{\LL}{\mbox{\boldmath$L$}}
\renewcommand{\SS}{\mbox{\boldmath$S$}}
\newcommand{\Tc}{$T_{\rm c}$ }
\newcommand{\Tcf}{$T_{\rm c}$}
\newcommand{\Co}{${\rm Na_{x}Co_{}O_{2}} \cdot y{\rm H}_{2}{\rm O}$ }
\newcommand{\Cof}{${\rm Na_{x}Co_{}O_{2}} \cdot y{\rm H}_{2}{\rm O}$}
\newcommand{\tgf}{$t_{\rm 2g}$-orbitals}
\newcommand{\tg}{$t_{\rm 2g}$-orbitals }
\newcommand{\av}{\mbox{\boldmath${\rm a}$} }
\newcommand{\bv}{\mbox{\boldmath${\rm b}$} }
\newcommand{\avf}{\mbox{\boldmath${\rm a}$}}
\newcommand{\bvf}{\mbox{\boldmath${\rm b}$}}
\newcommand{\egf}{$e_{\rm g}$-Fermi surface }
\newcommand{\egff}{$e_{\rm g}$-Fermi surface}
\newcommand{\agf}{$a_{\rm 1g}$-Fermi surface }
\newcommand{\agff}{$a_{\rm 1g}$-Fermi surface}

\section{Introduction}

 The pseudogap phenomena in high-\Tc cuprates have been 
one of the central issues in strongly correlated electron 
systems.~\cite{rf:timusk} 
 The theoretical elucidation of anomalous properties is highly desired 
for a comprehensive understanding of high-\Tc superconductivity.

 Among many theoretical proposals, the ``pairing scenario'' is 
one of the promising ideas. 
 Then, the origin of pseudogap is attributed to the superconducting 
correlation above \Tcf.~\cite{rf:janko,rf:yanasePG,rf:kobayashi,rf:metzner,
rf:yanasereview,rf:emery,rf:franz,rf:kwon,rf:loktev,rf:eckl,rf:tesanovic,
rf:herbut,rf:paramekanti} 
 In contrast to the conventional superconductors, the coherence length is 
extraordinary short in under-doped cuprates,  
which is typically a few lattice spacing. 
 The superconducting (SC) fluctuation is enhanced in such a short 
coherence length superconductor with quasi-2D lattice structure.

 In order to examine the validity of pairing scenario 
it is important to investigate the roles of disorder and/or 
magnetic field which suppress the $d$-wave superconductivity. 
 As for the magnetic field, it has been shown that the pseudogap 
induced by the SC fluctuation is slightly affected 
by the orbital de-pairing effect around the onset temperature 
$T=T^{*}$.~\cite{rf:yanaseMG} 
 This is mainly because the correlation length of superconductivity is
still short at $T=T^{*}$, where the pseudogap is induced by the 
short range SC correlation.
 On the other hand, the critical fluctuation around $T=T_{\rm c}$ is 
remarkably affected by the magnetic field because the correlation length 
diverges at the critical point.~\cite{rf:yanaseMG,rf:iyengar} 
 These results are consistent with NMR 
measurements.~\cite{rf:zheng,rf:gorny}

 In this paper, we investigate the role of disorder on the short 
coherence length superconductor with $d$-wave symmetry. 
 This issue is particularly interesting because the microscopic 
inhomogeneity observed in recent STM measurements~\cite{rf:pan,rf:mcelroy,
rf:sugimoto}
indicates a novel effect of disorder in the under-doped region. 
Furthermore, it is considered that the disorder plays an essential 
role in the superconductor-insulator (SI) transition in high-\Tc 
cuprates as implied by the disordered magnetism in the 
LSCO and YBCO compounds.~\cite{rf:keimer,rf:wakimoto,rf:bernhard,rf:sanna,
rf:ishidaPS}

 The pseudogap induced by the $d$-wave SC fluctuation has been 
investigated in disordered system on the basis of the self-consistent 
t-matrix approximation (SCTMA)~\cite{rf:comment} 
for the disorder.~\cite{rf:chen,rf:kudo} 
 Then, it was shown that the pseudogap and SC fluctuation 
are significantly {\it suppressed} by the disorder. 
 This conclusion seems to be reasonable since the \Tc in non-$s$-wave 
superconductors is suppressed by the disorder. Then, the 
thermal fluctuation can be suppressed.  
 However, this result is incompatible with the experimental 
results which have indicated a robustness of pseudogap in 
disorder-doped materials.~\cite{rf:itoh,rf:albenque,rf:xu} 

 In this paper, we carry out the calculation beyond the SCTMA where 
the disorder is exactly taken into account.  
 Contrary to the SCTMA, it is shown that the SC fluctuation is 
{\it enhanced} by the microscopic inhomogeneity which is characteristic 
in the quasi-2D short coherence length superconductor with 
non-$s$-wave symmetry. 
 Then, the long range coherence hardly develops owing to the 
nanoscale inhomogeneity of SC order parameter while the 
well developed short range correlation leads to the pseudogap in 
the low energy spectrum. 
 This result illuminates the breakdown of Abrikosov-Gorkov theory 
for the short coherence length superconductors.

 We analyze the $t$-$t'$-$V$ model to investigate the 
microscopic role of disorders on the SC fluctuation. 
 We adopt the $1$-loop order approximation with respect to the 
SC fluctuation, which is called ``self-consistent T-matrix 
approximation''.~\cite{rf:janko,rf:yanasePG,rf:kobayashi,rf:comment} 
 This approach is complementary with the phase 
fluctuation theory~\cite{rf:emery,rf:franz,rf:kwon,rf:loktev,
rf:eckl,rf:tesanovic,rf:herbut,rf:paramekanti} 
which is a phenomenological description for the deeply critical region. 
 It has been shown that the $1$-loop order theory is consistent 
with many experimental results in the pseudogap state.
 The anomalous behaviors in single particle, magnetic and transport 
properties~\cite{rf:ARPESreview,rf:timusk} are explained 
in a comprehensive way by taking into account the spin fluctuation 
in addition to the SC fluctuation.~\cite{rf:yanaseFLEXPG,rf:yanaseTRPG,
rf:yanasereview} 
 Furthermore, the microscopic estimation of the SC fluctuation 
in Hubbard model is consistent with the remarkable electron-hole 
asymmetry in high-\Tc cuprates.~\cite{rf:yanaseFLEXPG} 

 This paper is organized as follows. 
 The disordered $t$-$t'$-$V$ model is introduced in \S2, where 
we formulate the self-consistent T-matrix 
approximation~\cite{rf:comment} in disordered systems. 
 In \S3.1, the typical results on the spatial dependence of SC fluctuation 
are shown. In \S3.2, we numerically take the random average and 
calculate the DOS, SC correlation function and \Tcf. 
 The role of SC fluctuation is clarified by comparing with the 
mean field theory. 
 The effects of microscopic inhomogeneity are illuminated by the 
comparison with the SCTMA for disorder effects. 
 In \S4, we briefly discuss the spatial dependence of DOS around the 
single impurity. 
 Some discussions are given in the last section \S5.

\section{Formulation in the Disordered $t$-$t'$-$V$ Model}

\subsection{1-loop order theory in disordered systems}

 We adopt the disordered $t$-$t'$-$V$ model which is expressed as,  
\begin{eqnarray}
\label{eq:t-t'-V-model}
&& \hspace*{-10mm} H= -t \sum_{<i,j>,\sigma} c_{i\sigma}^{\dag}c_{j\sigma} 
+ t' \sum_{\ll i,j\gg,\sigma} c_{i\sigma}^{\dag}c_{j\sigma} 
\nonumber \\ && \hspace*{-5mm}
+ \frac{V}{2} \sum_{<i,j>} n_{i} n_{j}
+ \sum_{i} (W_{i}-\mu) n_{i}, 
\end{eqnarray}
where $n_{i}$ is the electron number at the site $i$. 
 The symbols $<i,j>$ and $\ll i,j\gg$ 
denote the summation over the nearest neighbor sites 
and that over the next nearest neighbor sites, respectively. 
 The attractive interaction $V<0$ between the nearest neighbor sites 
favors the $d_{\rm x^{2}-y^{2}}$-wave superconductivity around the 
half filling. 
 The disorder is introduced by the disordered potential $W_{i}$. 
 Although we have investigated several kinds of disorder 
including the box disorder, Gaussian disorder and $t'$-disorder, 
we show the results for the point disorder where $W_{i}=0$ or $W_{i}=W$. 
 Since many experimental studies on the Zn-doped cuprates have been 
reported, we assume the strong disorder $W=40t \gg t$. 
 Then, the sites $i$ where $W_{i}=W$ correspond to the Zn sites and 
the others are the Cu sites. 
 Recently, the disorder outside the plane has attracted much 
interests.~\cite{rf:mcelroy,rf:sugimoto,rf:fujita} 
 Then, the disordered potential is described by the extended 
point disorder. 
 We have confirmed that the following results are qualitatively 
independent of the type of disorder.

 It should be noted that the $t$-$t'$-$V$ model is a 
semi-phenomenological model for the short coherence length 
$d$-wave superconductor. 
 The $d$-wave pairing in high-\Tc cuprates is originally 
induced by the short range repulsive interaction through 
the many body effect. 
 We have analyzed the repulsive Hubbard model within the 
FLEX+T-matrix approximation,~\cite{rf:yanaseFLEXPG} 
and shown that the attractive models like 
eq.~(\ref{eq:t-t'-V-model}) appropriately capture the role 
of SC fluctuation.~\cite{rf:yanasereview} 
 In this paper, the effect of the microscopic inhomogeneity will 
be elucidated from the general point of view.

 We choose the unit of energy as $t=1$ and fix $t'/t=0.25$. 
 The chemical potential $\mu$ is fixed to $\mu=-0.8$ which corresponds
to the 10\% hole doping. 
 The typical Fermi surface of under-doped cuprates is 
well reproduced by this parameter set. 
 Since we fix the chemical potential instead of the total electron number, 
the average electron density per Cu site $n=\sum_{<i>} n_{i}/N_{0}$ 
is little affected by the disorder-doping. 
 Here, $\sum_{<i>}$ denotes the summation over 
the sites where $W_{\rm i}=0$. $N_{0}$ is the number of those sites. 
 Although the effect of impurity doping on the average electron density 
in high-\Tc cuprates is not clear, we can study the general roles of 
disorder by analyzing this model.

 We define the non-interacting part of Hamiltonian as follows, 
\begin{eqnarray}
\label{eq:H_0}
&& \hspace*{-8mm} H_{0}= -t \sum_{<i,j>,\sigma} c_{i\sigma}^{\dag}c_{j\sigma} 
+ t' \sum_{<<i,j>>,\sigma} c_{i\sigma}^{\dag}c_{j\sigma} 
+ \sum_{i} (W_{i}-\mu) n_{i}
\nonumber \\ &&
=
\sum_{\sigma} \hat{c}_{\sigma}^{\dag} \hat{H}_{0} \hat{c}_{\sigma}. 
\end{eqnarray}
Here, $\hat{c}_{\sigma}^{\dag}=(c_{i,\sigma}^{\dag})$ is the 
$N$-component vector and $N= L \times L$ is the total number of sites. 
 The Green function in the non-interacting system ($V=0$) is 
obtained by diagonalizing the $N \times N$ Hermitian matrix $\hat{H}_{0}$ 
as, 
\begin{eqnarray}
\label{eq:G_0}
G_{0}(\hat{i},\hat{j},\omega_{n}) = 
\sum_{l} u(m,l) 
\frac{1}{{\rm i}\omega_{n} - \varepsilon_{l}}
u^{*}(n,l), 
\end{eqnarray}
where $\varepsilon_{l}$ is the $l$-th eigenvalue of the matrix 
$\hat{H}_{0}$ and $u(m,l)$ is the $m$-th element of the eigenvector. 
 The index $m$ ($n$) is defined so that the $m$-th ($n$-th) element of 
$\hat{c}_{\sigma}^{\dag}$ is the creation operator 
at site $\hat{i}$ ($\hat{j}$). 
The fermion Matsubara frequency is given as $\omega_{n}=(2 n +1) \pi T$. 
 Since we have included the disorder potential into the unperturbed 
Hamiltonian, the disorder is exactly taken into account 
in the following calculation.

 In case of $V \ne 0$, the dressed Green function is obtained by the 
Dyson equation as, 
\begin{eqnarray}
\label{eq:G}
&& \hspace*{-15mm}  G(\hat{i},\hat{j},\omega_{n})= 
G_{0}(\hat{i},\hat{j},\omega_{n}) + 
\nonumber \\ &&\hspace*{5mm}
\sum_{\hat{i}',\hat{j}'}
G_{0}(\hat{i},\hat{i}',\omega_{n})
\Sigma(\hat{i}',\hat{j}',\omega_{n})
G(\hat{j}',\hat{j},\omega_{n}), 
\end{eqnarray}
where $\Sigma(\hat{i},\hat{j},\omega_{n})$ is the self-energy. 
 In this paper we estimate the self-energy within the 
self-consistent or non-self-consistent T-matrix approximation 
whose results in the clean system have been summarized in Ref.~14.

\begin{figure}[htbp]
  \begin{center}
\includegraphics[width=7cm]{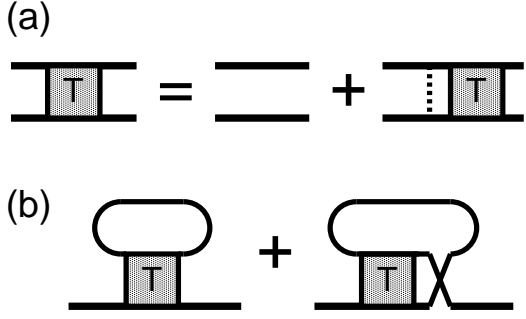}
\caption{The diagrammatic representation of (a) T-matrix and (b) 
         self-energy. 
             }
    \label{fig:diagrams}
  \end{center}
\end{figure}

 It generally takes a long computational time to calculate the 
self-energy in the inhomogeneous system because the Green function 
$G(\hat{i},\hat{j},\omega_{n})$ depends not only on the relative 
coordinate $\hat{i}-\hat{j}$ but also on the center of mass 
coordinate $(\hat{i}+\hat{j})/2$. 
 In order to reduce the computational time, we apply the quasi-static 
approximation for the propagator of SC fluctuation, namely the T-matrix. 
 The quantum dynamics of SC fluctuation, namely the imaginary time 
dependence of fluctuation propagator, is ignored in this approximation 
and only the thermal fluctuation is taken into 
account.~\cite{rf:leericeanderson,rf:sadovskii,rf:tchernyshyov1D,
rf:kopietzPG,rf:millisPG,rf:monienPG,rf:kuchinskii,rf:schmalianPG,
rf:tchernyshyov,rf:yanasePGVC}  
 This approximation is valid at finite temperature around 
the critical point. 
 We have used the quasi-static approximation for the estimation of 
higher order corrections beyond the $1$-loop order.~\cite{rf:yanasePGVC} 
 Then, we have found that this approximation is quantitatively 
justified in the pseudogap region. 
 It should be noted that the quasi-static approximation is particularly 
appropriate for the SC fluctuation owing to the characteristic dynamical 
property.~\cite{rf:yanasePGVC}

 Since the T-matrix having non-zero Matsubara frequency 
$\Omega_{n} \ne 0$ is ignored in the quasi-static approximation, 
we simply drop the variable of frequency in the T-matrix. 
 Then, the T-matrix is obtained by the following equations, 
\begin{eqnarray}
\label{eq:T-matrix}
&& \hspace*{-12mm} T(\hat{i},\hat{j},\hat{\delta},\hat{\delta}') = 
T_{0}(\hat{i},\hat{j},\hat{\delta},\hat{\delta}') - 
\nonumber \\ && \hspace*{9mm}
\sum_{k,\hat{\delta}''}
T_{0}(\hat{i},\hat{k},\hat{\delta},\hat{\delta}'') V 
T(\hat{k},\hat{j},\hat{\delta}'',\hat{\delta}'), 
\\
\label{eq:bare-T-matrix}
&&  \hspace*{-12mm} 
T_{0}(\hat{i},\hat{j},\hat{\delta},\hat{\delta}')
=T \sum_{m} G(\hat{i},\hat{j},\omega_{m}) 
G(\hat{i}+\hat{\delta},\hat{j}+\hat{\delta}',-\omega_{m}), 
\end{eqnarray}
where $\hat{\delta}=(\pm 1,0)$ or $\hat{\delta}=(0,\pm 1)$. 
 The diagrammatic representation of T-matrix is shown in Fig.~1(a).

 The self-energy is represented in Fig.~1(b) and the expression is 
given as, 
\begin{eqnarray}
\label{eq:self-energy}
&& \hspace*{-10mm} \Sigma(\hat{i},\hat{j},\omega_{n}) = 
-V^{2}
\sum_{\delta,\delta'} 
[T_{\rm s}(\hat{i},\hat{j},\hat{\delta},\hat{\delta}')
+3 T_{\rm t}(\hat{i},\hat{j},\hat{\delta},\hat{\delta}')
]
\nonumber \\ && \hspace*{20mm} \times
G(\hat{j}+\hat{\delta}',\hat{i}+\hat{\delta},-\omega_{n}), 
\end{eqnarray}
where 
\begin{eqnarray}
\label{eq:singlet-triplet}
&& \hspace*{-12mm}
T_{\rm s,t}(\hat{i},\hat{j},\hat{\delta},\hat{\delta}') 
= \frac{1}{2} T(\hat{i},\hat{j},\hat{\delta},\hat{\delta}') 
\pm \frac{1}{2} T(\hat{i},\hat{j}+\hat{\delta}',\hat{\delta},-\hat{\delta}').
\end{eqnarray}
 Here, $T_{\rm s}(\hat{i},\hat{j},\hat{\delta},\hat{\delta}')$ 
($T_{\rm t}(\hat{i},\hat{j},\hat{\delta},\hat{\delta}')$) 
is the fluctuation propagator for the spin singlet (triplet) pairing. 
 The former includes the $d_{\rm x^{2}-y^{2}}$-wave and extended 
$s$-wave pairings while the latter includes the 
$p_{\rm x}$-wave and $p_{\rm y}$-wave pairings. 
 The coefficient $3$ of 
$T_{\rm t}(\hat{i},\hat{j},\hat{\delta},\hat{\delta}')$  in 
eq.~(\ref{eq:self-energy}) is due to the spin $1$ of Cooper pairs 
in the spin triplet pairing. 

 Although the $d$-wave pairing state is most stable 
in the $t$-$t'$-$V$ model, the pairing correlation develops also 
in the $p$-wave channel. 
 The latter is expected to be an artifact of the $t$-$t'$-$V$ model 
because the $p$-wave pairing instability is negligible in the 
microscopic model, such as the Hubbard, $d$-$p$ and $t$-$J$ 
models.~\cite{rf:yanasereview,rf:dagottoreview,rf:bulutreview} 
 Therefore, we simply drop the contribution from the 
spin triplet SC fluctuation in eq.~(\ref{eq:self-energy}) and 
obtain the self-energy as, 
\begin{eqnarray}
\label{eq:self-energy-singlet}
&& \hspace*{-14mm}
\Sigma(\hat{i},\hat{j},\omega_{n}) = 
-V^{2}
\sum_{\delta,\delta'} 
T_{\rm s}(\hat{i},\hat{j},\hat{\delta},\hat{\delta}')
G(\hat{j}+\hat{\delta}',\hat{i}+\hat{\delta},-\omega_{n}). 
\end{eqnarray}
 It should be again stressed that the $t$-$t'$-$V$ model is a 
semi-phenomenological model which is relevant for the $d$-wave 
SC fluctuation. 
 Note that the extended $s$-wave pairing fluctuation is 
also included in 
$T_{\rm s}(\hat{i},\hat{j},\hat{\delta},\hat{\delta}')$ 
and we cannot separate these pairings in the disordered system 
owing to the violation of $4$-fold rotational symmetry. We have 
confirmed that the contribution from the extended $s$-wave pairing 
is not important.

 In the self-consistent T-matrix approximation, the self-energy, 
Green function and T-matrix are obtained by solving the self-consistent
equations, namely eqs.~(\ref{eq:G_0}-\ref{eq:bare-T-matrix}) and  
(\ref{eq:singlet-triplet}-\ref{eq:self-energy-singlet}). 
 The non-self-consistent T-matrix approximation is given by replacing 
the Green function $G$ in eqs.~(\ref{eq:bare-T-matrix}) and 
(\ref{eq:self-energy-singlet}) with the unperturbed one $G_0$. 
 The self-consistent calculation is carried out for up to 
$N=31 \times 31$ sites, while the non-self-consistent one is performed 
for up to $N=65 \times 65$ sites.

\subsection{SCTMA for disorder}

 The superconductivity in the disordered alloy has been investigated 
for more than four decades.~\cite{rf:abrikosov} 
 The Born approximation or SCTMA have been used to discuss 
the macroscopic properties. 
 Then, the random average is taken in advance and the translational 
symmetry is reestablished in the results. 
 The role of strong disorder on the gap-less superconductivity has been 
elucidated with use of the SCTMA.~\cite{rf:maki,rf:hotta,rf:hirschfeld} 
 However, the role of inhomogeneity is not appropriately taken into 
account in the SCTMA because the interference effect of many disorders 
is neglected. 
 In contrast to that, the disorder is exactly taken into account 
in the formulation in \S2.1. 
 Therefore, it is interesting to illuminate the role of microscopic 
inhomogeneity by the comparison with SCTMA. 
 In this subsection, we briefly explain the formulation of SCTMA in 
the $t$-$t'$-$V$ model.

 Since the disorder is taken into account in an approximate way, 
the unperturbed Hamiltonian is defined as, 
\begin{eqnarray}
\label{eq:H'_0}
H'_{0}=- t \sum_{<i,j>,\sigma} c_{i\sigma}^{\dag}c_{j\sigma} 
+ t' \sum_{<<i,j>>,\sigma} c_{i\sigma}^{\dag}c_{j\sigma} 
- \mu \sum_{i,\sigma} n_{i}. 
\end{eqnarray}
 Then, the Green function, self-energy and T-matrix can be written in 
the $k$-space representation. 
 For example, the Green function is written as,~\cite{rf:kudo,rf:chen}
\begin{eqnarray}
\label{eq:SCTMA_G}
G(\k,\omega_{n}) = 
\frac{1}{{\rm i}\omega_{n} - \varepsilon(\k) 
- \Sigma_{\rm s}(\k,\omega_{n}) - \Sigma_{\rm d}(\omega_{n})}, 
\end{eqnarray}
where $\varepsilon(\k)$ is the dispersion relation, 
\begin{eqnarray}
\label{eq:high-tc-dispersion}
\varepsilon(\k)=-2t(\cos k_{\rm x}+\cos k_{\rm y})
+4t'\cos k_{\rm x} \cos k_{\rm y}-\mu. 
\end{eqnarray}

 The self-energy is given by the contribution from the 
SC fluctuation $\Sigma_{\rm s}(\k,\omega_{n})$ and that from the 
disorder $\Sigma_{\rm d}(\omega_{n})$. 
 In order to compare with the formulation in \S2.1, 
we calculate the self-energy $\Sigma_{\rm s}(\k,\omega_{n})$ 
in the same approximation as in \S2.1. 
 Then, $\Sigma_{\rm s}(\k,\omega_{n})$ is expressed in the real space 
as, 
\begin{eqnarray}
\label{eq:SCTMA_self-energy_s}
\Sigma_{\rm s}(\hat{i},\omega_{n}) = 
-V^{2}
\sum_{\delta,\delta'} 
T_{\rm s}(\hat{i},\hat{\delta},\hat{\delta}')
G(-\hat{i}-\hat{\delta}+\hat{\delta}',-\omega_{n}), 
\end{eqnarray}
where 
\begin{eqnarray}
\label{eq:SCTMA_singlet-triplet}
T_{\rm s}(\hat{i},\hat{\delta},\hat{\delta}') 
= \frac{1}{2} T(\hat{i},\hat{\delta},\hat{\delta}') 
+ \frac{1}{2} T(\hat{i}-\hat{\delta}',\hat{\delta},-\hat{\delta}'),
\end{eqnarray}
and
\begin{eqnarray}
\label{eq:SCTMA_T-matrix}
&& \hspace*{-10mm}  T(\hat{i},\hat{\delta},\hat{\delta}') = 
T_{0}(\hat{i},\hat{\delta},\hat{\delta}') - 
\nonumber \\ && \hspace*{7mm}
\sum_{k,\hat{\delta}''}
T_{0}(\hat{i}-\hat{k},\hat{\delta},\hat{\delta}'') V 
T(\hat{k},\hat{\delta}'',\hat{\delta}'), 
\\
\label{eq:SCTMA_bare-T-matrix}
&& \hspace*{-10mm}
T_{0}(\hat{i},\hat{\delta},\hat{\delta}')
=T \sum_{m} G(\hat{i},\omega_{m}) 
G(\hat{i}+\hat{\delta}-\hat{\delta}',-\omega_{m}). 
\end{eqnarray}
 The $k$-space representations of $\Sigma_{\rm s}(\k,\omega_{n})$ and 
$G(\k,\omega_{n})$ are obtained from the Fourier transformation of 
$\Sigma_{\rm s}(\hat{i},\omega_{n})$ and $G(\hat{i},\omega_{n})$, 
respectively.

\begin{figure}[htbp]
  \begin{center}
\includegraphics[width=3.5cm]{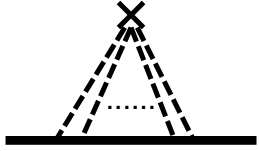}
\caption{Self-energy arising from the impurity scattering in the SCTMA.
             }
    \label{fig:SCTMA}
  \end{center}
\end{figure}

 The self-energy arising from the disorder is given as, 
\begin{eqnarray}
\label{eq:SCTMA_self-energy_d}
\Sigma_{\rm d}(\omega_{n}) = 
- \frac{n_{\rm imp}}{\sum_{k} G(\k,\omega_{n})}, 
\end{eqnarray}
where $n_{\rm imp}=(N-N_0)/N$ is the impurity concentration. 
 The diagrammatic representation of eq.~(\ref{eq:SCTMA_self-energy_d}) 
is shown in Fig.~2.  
 Here, we have taken the limit $W \rightarrow \infty$ for simplicity, 
which is consistent with the sufficiently strong potential $W=40 t$ 
in \S2.1. 
 We have ignored the vertex correction arising from the disorder on the 
T-matrix because it is negligible for the $d$-wave SC 
fluctuation.~\cite{rf:kudo,rf:chen}

 If we neglect the self-energy due to the SC fluctuation 
$\Sigma_{\rm s}(\k,\omega_{n})$, this calculation is reduced to the 
SCTMA in the mean field theory. 
 The deviation from the pure $d$-wave superconductivity in the 
experimental results, such as the NMR $1/T_{1}T$ and magnetic field 
penetration depth, has been resolved within the mean field 
theory.~\cite{rf:hotta,rf:maki,rf:hirschfeld} 
 The consistency between the theory and experiment has been regarded as 
an evidence for the $d$-wave pairing in high-\Tc 
cuprates.~\cite{rf:ishida,rf:hardy} 
 We show that this approximation breaks down in the under-doped region, 
while the SCTMA will be valid in the over-doped and electron-doped 
region.

\section{Microscopic Inhomogeneity and SC Fluctuation}

\subsection{Microscopic inhomogeneity}

 First of all, we show the typical spatial dependence of SC order
parameter in the disordered short coherence length superconductor. 
 In order to discuss the long range correlation, we show the results of 
non-self-consistent T-matrix approximation for $N=65 \times 65$ sites 
in this subsection. 
 We have confirmed that qualitatively the same results are obtained in 
the self-consistent calculation.

 We determine the spatial dependence of SC order parameter 
from the fluctuation propagator, namely the T-matrix. 
 Although the amplitude of SC order parameter is not determined by the 
T-matrix, the spatial dependence is determined in the following way. 
 First, we solve the eigenvalue equation to diagonalize the 
irreducible T-matrix as, 
\begin{eqnarray}
\label{eq:irreducibleT-matrix} 
t_{n} v(n,\hat{i},\hat{\delta}) = 
T_{0}(\hat{i},\hat{j},\hat{\delta},\hat{\delta}') 
v(n,\hat{j},\hat{\delta}'). 
\end{eqnarray}
 Here, $t_n$ is the eigenvalue of 
$T_{0}(\hat{i},\hat{j},\hat{\delta},\hat{\delta}')$ 
and we define $t_1 \geq t_2 \geq t_3 ...... \geq t_{4N}$. 
 Then, the T-matrix is described as, 
\begin{eqnarray}
\label{eq:diagonalizeT-matrix} 
T(\hat{i},\hat{j},\hat{\delta},\hat{\delta}') 
=\sum_{n} v(n,\hat{i},\hat{\delta})  \frac{t_{n}}{1 + V t_{n}} 
v(n,\hat{j},\hat{\delta}'), 
\end{eqnarray}
and the critical temperature is determined by the criterion $|V| t_1 = 1$. 
The order parameter of $d$-wave superconductivity just below \Tc is 
obtained as, 
\begin{eqnarray}
\label{eq:dSCOP} 
\Delta_{\rm d}(\hat{i})
=\sum_{\delta} (-1)^{P} 
v(1,\hat{i},\hat{\delta}), 
\end{eqnarray}
where $P=0$ ($P=1$) for $\hat{\delta}=(\pm 1,0)$ 
($\hat{\delta}=(0,\pm 1)$). 
 The extended $s$-wave component of the order parameter is also 
determined as, 
\begin{eqnarray}
\label{eq:sSCOP} 
\Delta_{\rm s}(\hat{i})
=\sum_{\delta} v(1,\hat{i},\hat{\delta}). 
\end{eqnarray}
 Although $\Delta_{\rm s}(\hat{i})$ is finite in the disordered system, 
it is much smaller than the $d$-wave component $\Delta_{\rm d}(\hat{i})$. 
 The small $s$-wave component does not play any important role in the 
following results.

 We show the typical spatial dependence of $\Delta_{\rm d}(\hat{i})$ 
in Fig.~3. 
 We choose the parameter as $V=-0.6$ in Fig.~3(a) while 
we choose $V=-1.5$ in Figs.~3(b) and (c). 
 The disorder concentration is chosen to be $n_{\rm imp}=0.01$ 
in Figs.~3(a) and (b) and $n_{\rm imp}=0.05$ in Fig.~3(c). 
 Note that the coherence length in the clean limit $\xi_{0}$ is 
scaled by $T_{\rm c}^{0}$ as $\xi_{0} \sim 
v_{\rm F}/T_{\rm c}^{0}$ where $T_{\rm c}^{0}$ is the 
transition temperature in the clean limit and $v_{\rm F}$ is the 
Fermi velocity. 
 We obtain $T_{\rm c}^{0}=0.0418$ and $T_{\rm c}^{0}=0.269$ for 
$V=-0.6$ and $V=-1.5$, respectively. 
 Thus, the coherence length in Fig.~3(a) is much longer than 
that in Figs.~3(b) and (c).

\begin{figure}[htbp]
  \begin{center}
\hspace*{-10mm}
\includegraphics[width=10cm]{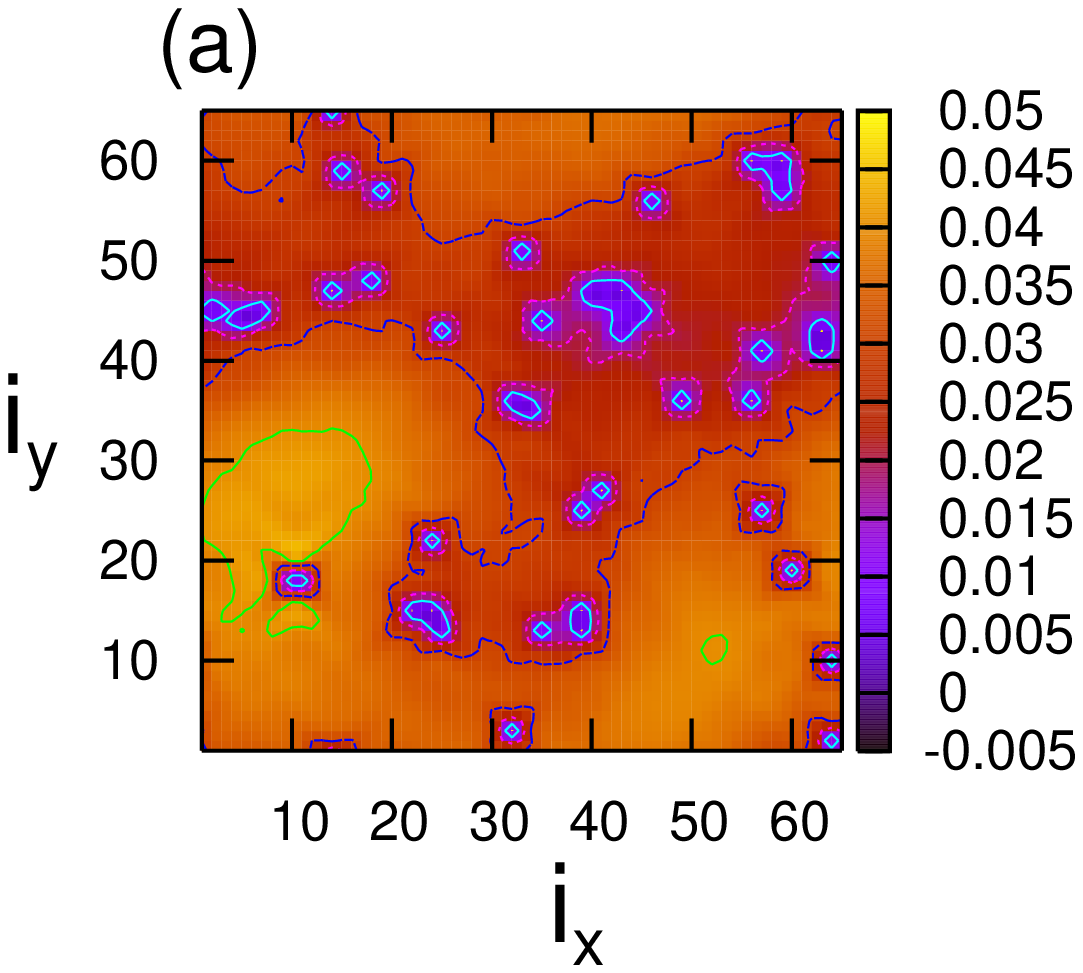}
\hspace*{-10mm}
\includegraphics[width=10cm]{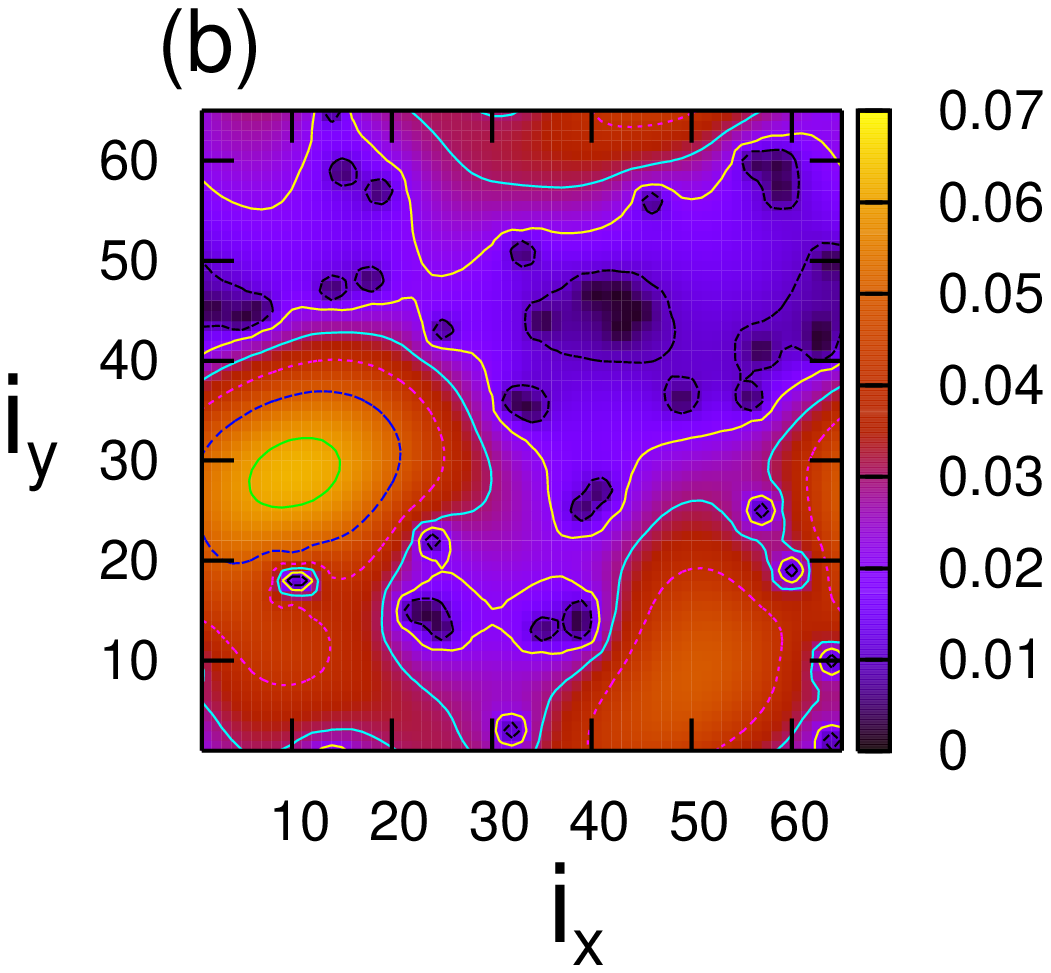}
\hspace*{-10mm}
\includegraphics[width=10cm]{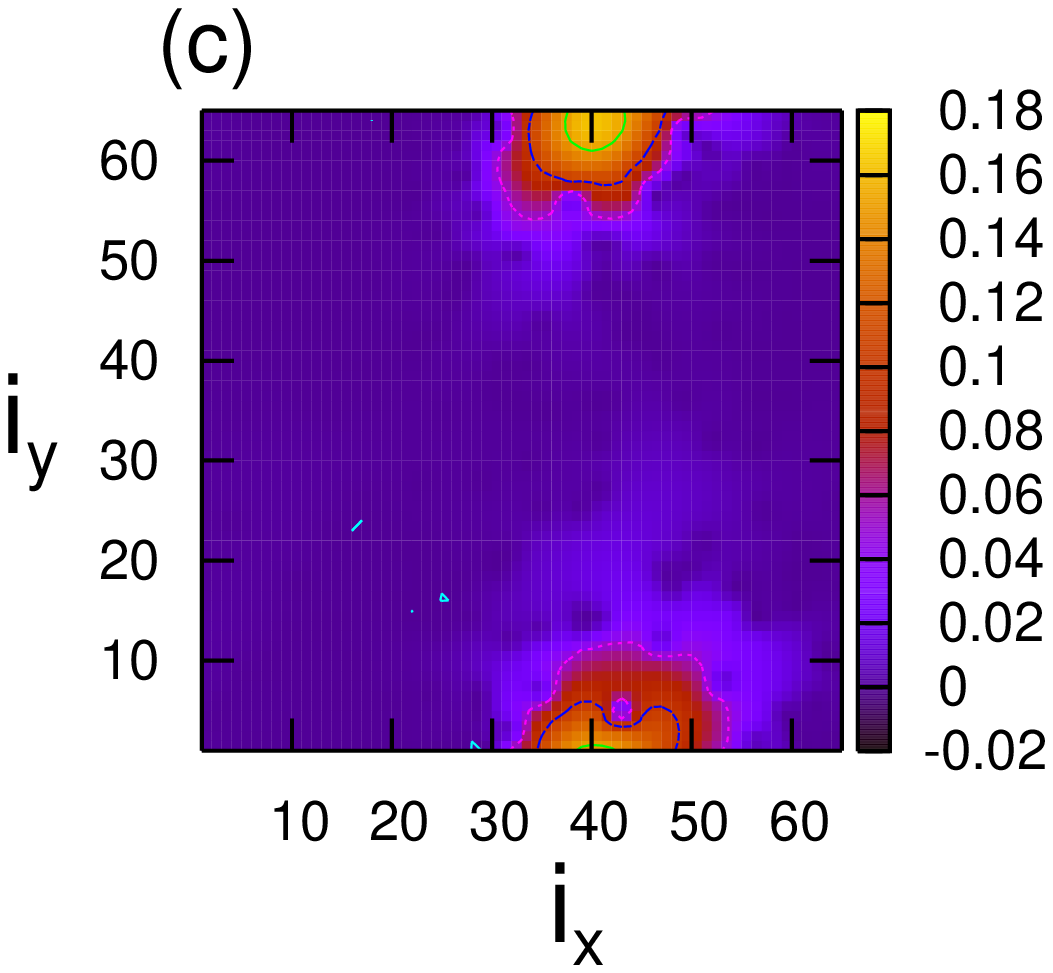}
\caption{Spatial dependence of $d$-wave SC order parameter 
         at $T=T_{\rm c}^{\rm MF}$. 
         (a) The case of long coherence length superconductor,  
             $V=-0.6$ and $n_{\rm imp}=0.01$. 
         (b) The case of short coherence length superconductor,  
             $V=-1.5$ and $n_{\rm imp}=0.01$. 
         (c) $V=-1.5$ and $n_{\rm imp}=0.05$. 
             }
    \label{fig:dSCOP}
  \end{center}
\end{figure}

 We clearly see the remarkable spatial inhomogeneity of SC order 
parameter in Figs.~3(b) and (c). 
 In particular, the SC order parameter is highly localized by the 5\% 
disorder-doping in the case of $V=-1.5$. 
 It should be noticed that these results are very unusual.  
 In case of the conventional long coherence length superconductor, 
the SC order parameter is almost uniform in the microscopic scale as 
shown in Fig.~3(a). 
 If we increase the disorder with fixing $V=0.6$, the superconductivity 
is depressed before the SC order parameter is localized. 
 This is the usual behavior of non-$s$-wave superconductor where 
the superconductivity is suppressed by the non-magnetic impurities. 
 Then, the inhomogeneity is expected only in the larger scale.

 We understand these results by considering the two typical scales 
of impurity concentration. 
 One is the critical concentration defined by the Abrikosov-Gorkov 
theory.~\cite{rf:abrikosov,rf:maki,rf:hotta,rf:hirschfeld} 
 The other is defined so that the mean length between impurities 
becomes comparable with the coherence length in the clean limit, 
namely $\xi_{0}$. 
 The former is scaled as 
$n_{\rm imp} \propto T_{\rm c}^{0}/E_{\rm F}$ where $E_{\rm F}$ 
is the Fermi energy. 
 The latter is obtained as 
$n_{\rm imp} \sim \xi_{0}^{-2} \propto (T_{\rm c}^{0}/E_{\rm F})^{2}$ 
in two dimension. 
 The latter is much smaller than the former in the low temperature 
superconductors, and therefore the system has many impurities in 
the area $\sim \xi_{0}^{2}$ at the moderate impurity concentration. 
This is the underlying reason why the Abrikosov-Gorkov theory is justified.

\begin{figure}[htbp]
  \begin{center}
\includegraphics[width=5.3cm]{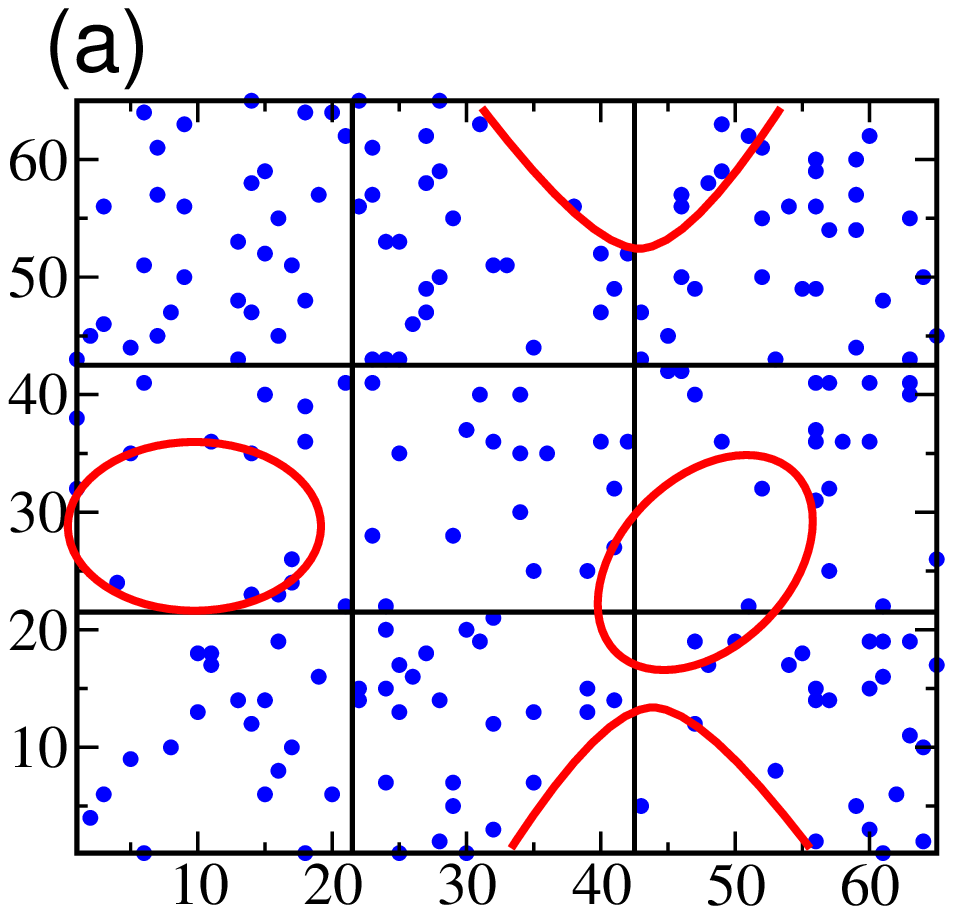}
\hspace*{-10mm}
\includegraphics[width=10cm]{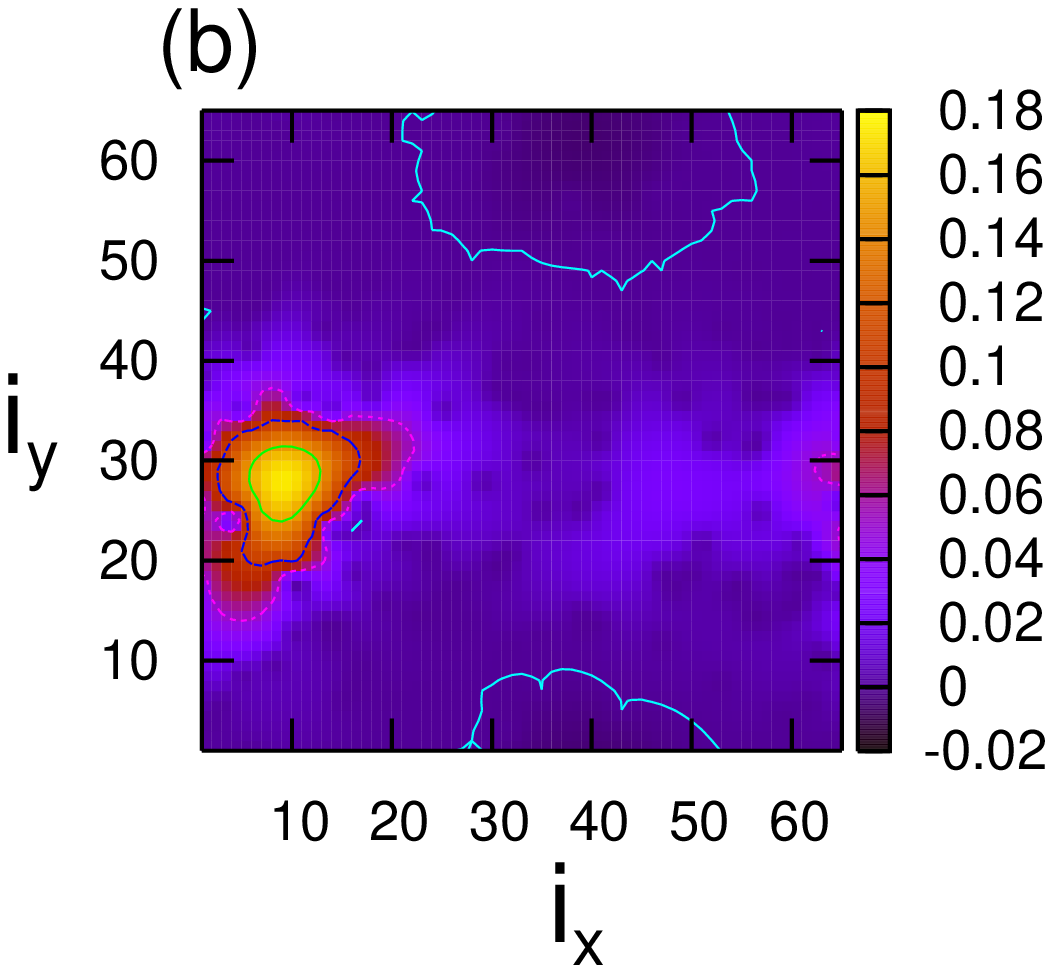}
\hspace*{-10mm}
\includegraphics[width=10cm]{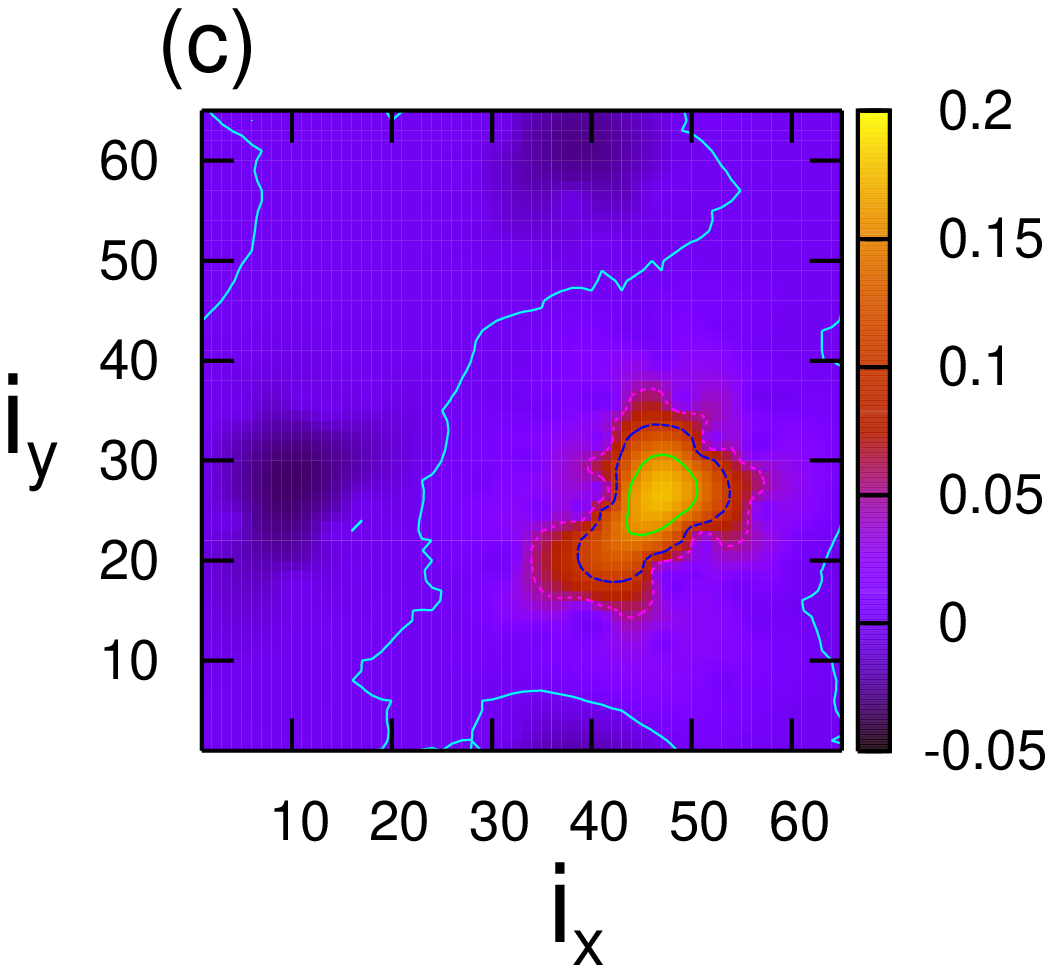}
\caption{(a) The distribution of point disorders in Fig.~3(c). 
             We show the ``clean region'' by red circles. 
             The SC correlation is locally enhanced in these regions. 
         (b) Spatial dependence of $\Delta_{2}(\hat{i})$ 
         and (c) that of $\Delta_{3}(\hat{i})$. 
         We choose the same parameters as in Fig.~3(c).
             }
    \label{fig:pdotcircle}
  \end{center}
\end{figure}

 In contrast to the conventional case, the two scales can be comparable 
in the short coherence length superconductor, such as the high-\Tc 
and organic superconductors. Then, the local superconductivity is 
stabilized in the ``clean region'' where only a few impurities exist. 
 For example, we show the distribution of point disorders for 
$n_{\rm imp}=0.05$ in Fig.~4(a).  
 Comparing with Fig.~3(c), we see that the SC order 
parameter is enhanced in the nanoscale ``clean region'' which
is shown by the red circles. 
 The SC order parameter in Fig.~3(c) is localized in one of 
the red circles in Fig.~4(a). 
 We understand that the SC correlation develops in the other 
``clean regions'' by showing 
$
\Delta_{2}(\hat{i})
=\sum_{\delta} (-1)^{P} 
v(2,\hat{i},\hat{\delta}) 
$
and 
$
\Delta_{3}(\hat{i})
=\sum_{\delta} (-1)^{P} 
v(3,\hat{i},\hat{\delta}) 
$
which are the $d$-wave order parameters having the second and third 
highest \Tcf, respectively. 
 It is shown in Figs.~\ref{fig:pdotcircle}(b) and (c) 
that they are developed in one of the red circles in Fig.~4(a). 
 Thus, the highly disordered $d$-wave superconductor with short 
coherence length has a granular-like spatial structure. 
 It should be stressed that the length scale of each granular is 
nanometer scale which is much smaller than the conventional 
granular systems with macroscopic or mesoscopic scale.

 The microscopic inhomogeneity in Figs.~3(b), 3(c) and 4 is consistent 
with the STM measurement in the under-doped region.~\cite{rf:pan} 
 This experimental observation is the fingerprint of the 
$d$-wave short coherence length superconductor with point-like disorders. 
 This interpretation is consistent with the recent experimental 
report~\cite{rf:mcelroy} where the spatial structure of 
SC order parameter is associated with the extended but point 
impurities. 
 Then, the clean region is characterized by the small gap and the 
sharp gap edge.~\cite{rf:hasegawa} 
The broad gap structure in the dirty region should be attributed 
to the competing order.~\cite{rf:yanasetUV,rf:hirschfeldmf}

\begin{figure}[htbp]
  \begin{center}
\includegraphics[width=7.5cm]{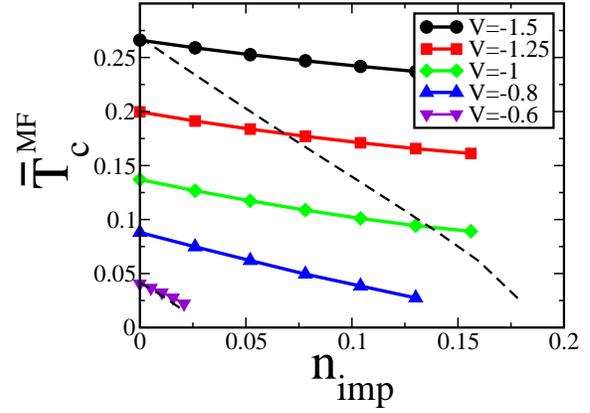}
\caption{Random average of the transition temperature 
         $\bar{T}_{\rm c}^{\rm MF}$ in the mean field theory. 
         Here, we take more than $50$ samples for the random average. 
         The attractive interaction is varied as $V=-0.6$, $-0.8$, 
         $-1$, $-1.25$ and $-1.5$ from the bottom to the top.  
         We show the results of SCTMA with $V=-0.6$ and $V=-1.5$ 
         (dashed lines) for a comparison. 
             }
    \label{fig:meanfieldTc}
  \end{center}
\end{figure}

 As a result of the nanoscale inhomogeneity of SC order parameter, 
the mean field value of transition temperature $T_{\rm c}^{\rm MF}$ 
is robust for the disorder, as pointed out in Ref.~58. 
 This is simply because the localized order parameter in the 
``clean region'' is little affected by the disorder potential. 
 We numerically take the random average of transition temperature 
$\bar{T}_{\rm c}^{\rm MF}$ and show the result in Fig.~5. 
 The result of SCTMA is also shown for a comparison. 
 The decrease of the mean field transition temperature remarkably 
deviates from the SCTMA in case of the large $|V|$, namely the 
short coherence length ($V=-1.5$) while the deviation is negligible 
in case of the long coherence length ($V=-0.6$). 
 Thus, the Abrikosov-Gorkov theory breaks down in the presence of 
the nanoscale inhomogeneity.

 Before closing this subsection, we note that the spatial dependence 
of SC order parameter is quite different in the $s$-wave superconductor. 
 Then, the order parameter is almost uniform for 10\% disorder-doping 
even if the coherence length is short.~\cite{rf:yanaseswave} 
 This is because the $s$-wave superconductivity is robust for the 
non-magnetic impurity as proved by Anderson.~\cite{rf:andersontheorem} 
 We obtain the spatial dependence similar to Fig.~3(a) where 
the SC order parameter is nearly homogeneous except for the 
abrupt suppression around impurities. 
 Thus, the $d$-wave pairing symmetry plays an essential role in the 
inhomogeneous structure of SC order parameter in the nanometer scale. 
 This may be the reason why the critical behavior in short coherence 
length $s$-wave superconductor NbN is quite different from that in 
the under-doped cuprates.~\cite{rf:kitano}

\subsection{SC fluctuation in the inhomogeneous system}

 The purpose of this paper is to investigate the SC fluctuation 
in the disordered system with focus on the role of microscopic 
inhomogeneity. 
 We discuss this issue by showing the results of self-consistent 
T-matrix approximation.

 First, we discuss the transition temperature of superconductivity. 
 Because the long range order is not realized at finite temperature 
in purely two-dimensional systems according to the Mermin-Wagner theorem, 
we adopt a phenomenological procedure to include the weak three 
dimensionality. For this aim, the criterion $|V|t_{1}=1-\delta$ 
has been adopted instead of $|V|t_{1}=1$ when the clean system 
has been investigated.~\cite{rf:yanasereview}
 We also adopt this procedure and determine the \Tc for 
$n_{\rm imp}=0$ by choosing $\delta=0.02$. 
 However, it is not clear whether this criterion is applicable to 
the disordered systems. In particular, this procedure may break down 
in the disordered short coherence length superconductor with 
nanoscale inhomogeneity. 
 Then, it is expected that the long range correlation hardly develops 
although the local superconductivity occurs as in Figs.~3 and 4. 
 Therefore, we determine the \Tc by the following procedure. 
 First, we calculate the correlation function of $d$-wave SC order 
parameter $\bar{\chi}_{\rm d}(\hat{r})$, which is defined as, 
\begin{eqnarray}
\chi_{\rm d}(\hat{i},\hat{j}) = 
\sum_{\delta,\delta'} (-1)^{P+P'} 
T(\hat{i},\hat{j},\hat{\delta},\hat{\delta}'), 
\label{eq:local-correlation-function} 
\\
\bar{\chi}_{\rm d}(\hat{r})=
<\frac{1}{N}\sum_{i}\chi_{\rm d}(\hat{i}+\hat{r},\hat{i})>_{\rm r}, 
\label{eq:correlation-function} 
\end{eqnarray}
where $<>_{\rm r}$ denotes the random average. 
 Here, $P$ ($P'$) is determined by $\hat{\delta}$ 
($\hat{\delta}'$) as in eq.~(\ref{eq:dSCOP}). 
 We denote the correlation function for $n_{\rm imp}=0$ and 
$|V|t_{1}=0.98$ as $\chi_{\rm d}^{\rm (c)}(\hat{r})$ and 
adopt the criterion, 
\begin{eqnarray}
\bar{\chi}_{\rm d}(15,15)=\chi_{\rm d}^{\rm (c)}(15,15), 
\label{eq:Tc-criterion} 
\end{eqnarray}
to determine the \Tc in disordered systems. 
 Note that $\hat{r}=(15,15)$ is the longest length in the calculation 
for $N=31 \times 31$ with (anti-)periodical boundary condition. 
 This criterion determines the temperature where the 
long range correlation begins to develop.

\begin{figure}[htbp]
  \begin{center}
\includegraphics[width=7.5cm]{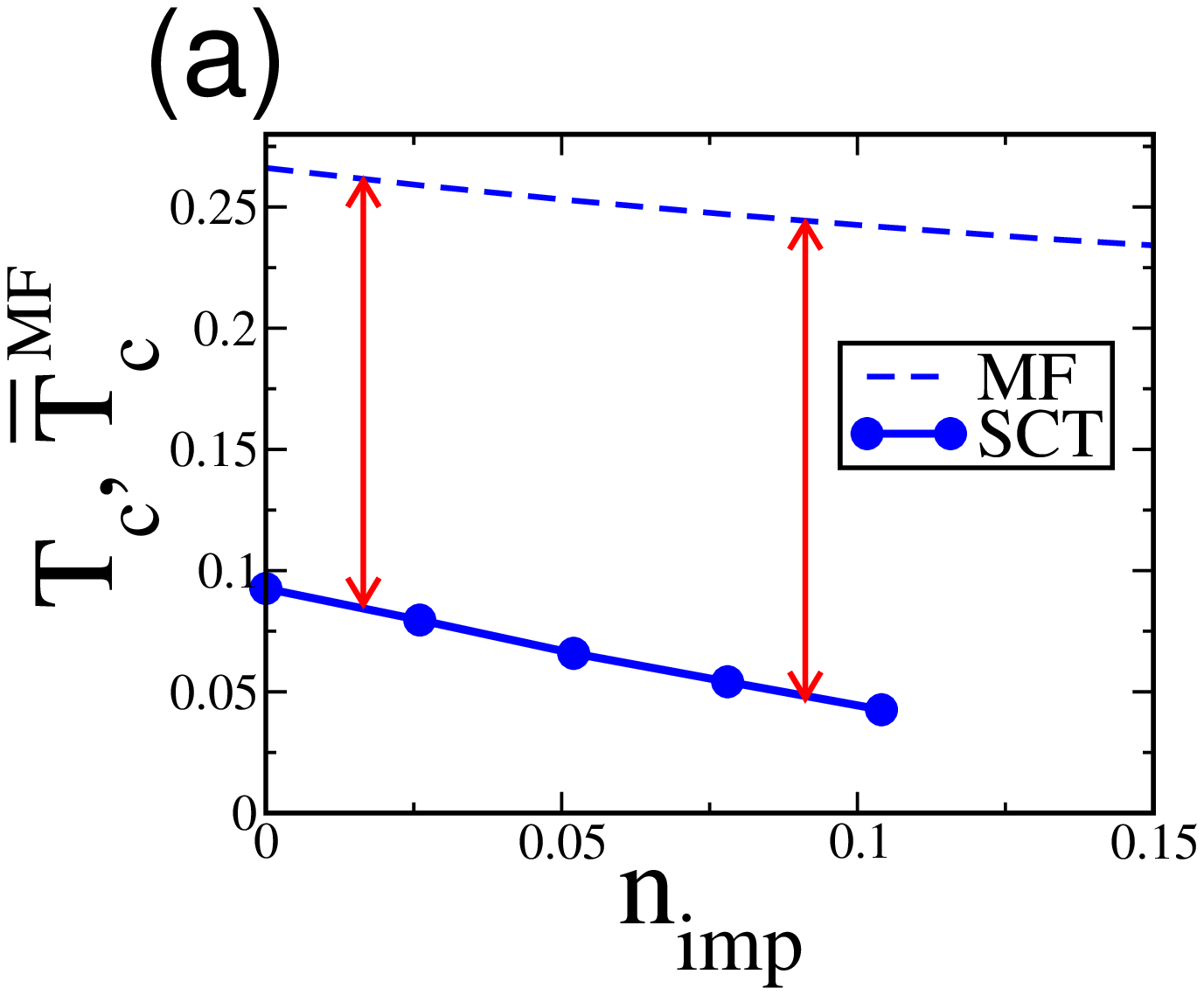}
\hspace{3mm}
\includegraphics[width=7.5cm]{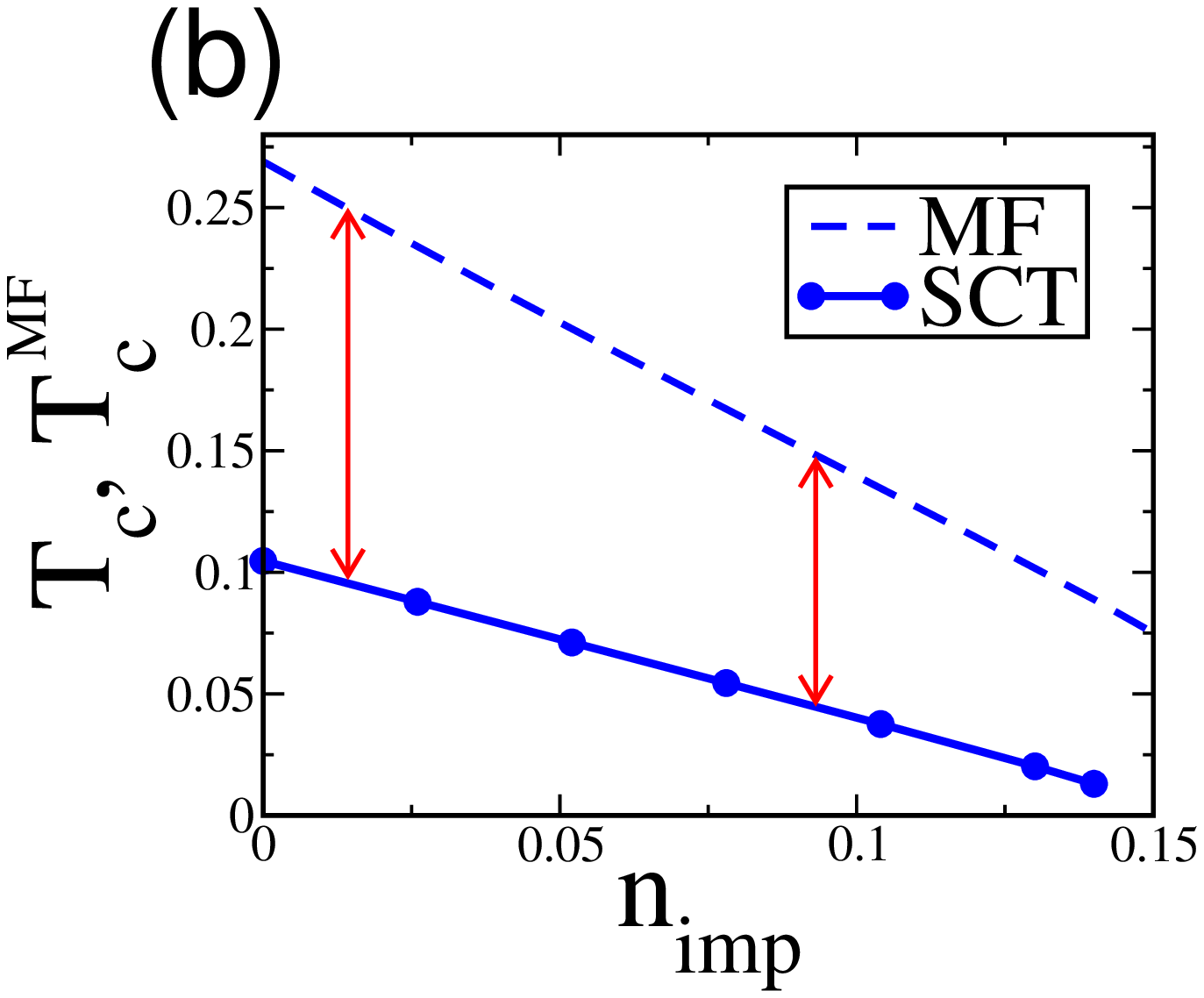}
\caption{
The transition temperature obtained in the self-consistent 
T-matrix approximation for $V=-1.5$. 
(a) The results of the formulation in \S2.1 where the disorder is exactly 
taken into account. We take at least 20 samples and take the random 
average to estimate the correlation function, 
eq.~(\ref{eq:correlation-function}). 
(b) The results of SCTMA. 
We show the transition temperature in the mean field theory, 
$\bar{T}_{\rm c}^{\rm MF}$ and $T_{\rm c}^{\rm MF}$ by the dashed lines. 
             }
    \label{fig:rspacefltc}
  \end{center}
\end{figure}

 We show the results of \Tc in Fig.~6(a). 
 The results obtained in the SCTMA are shown in Fig.~6(b) for a comparison. 
 The same criterion, namely eq.~(\ref{eq:Tc-criterion}) is adopted 
in the SCTMA. 
 The transition temperatures in the mean field theory 
((a) $\bar{T}_{\rm c}^{\rm MF}$ and (b) $T_{\rm c}^{\rm MF}$, 
respectively) are shown to illuminate the role of SC fluctuation. 

 Since the short range SC fluctuation develops below 
the mean field transition temperature, 
$\bar{T}_{\rm c}^{\rm MF}$ can be regarded as $T^{*}$ which 
is the onset of pseudogap phenomena arising from the SC fluctuation. 
 An important value is the difference 
$\bar{T}_{\rm c}^{\rm MF}-T_{\rm c}$ which measures the width of 
critical region. 
 When the SC fluctuation is strong, the critical region is large 
and {\it vice versa}.

 We clearly see the qualitatively different behaviors of critical 
region between Figs.~6(a) and (b). 
 The results of the formulation in \S2.1 show that the critical region 
is {\it enlarged} by the disorder. However, the SCTMA provides an 
opposite result. 
 This discrepancy means that the result of Fig.~6(b) is an artifact 
of the SCTMA because the disorder is exactly taken into account in 
Fig.~6(a). This comparison illuminates the breakdown of Abrikosov-Gorkov 
theory for the fluctuation phenomena in the short coherence 
superconductor.~\cite{rf:andoAGbreak}

 The result of SCTMA, where the SC fluctuation is {\it suppressed} 
by the disorder, is seemingly reasonable because the \Tc is 
decreased by the disorder. 
 However, the microscopic inhomogeneity neglected in the SCTMA 
{\it enhances} the fluctuation and disturbs the long range order. 
 The latter effect is significant in the short coherence length 
superconductor like under-doped cuprates and organic superconductor 
$\kappa$-(BEDT-TTF)$_2$X. 
 Thus, the {\it enhancement} of SC fluctuation indicated by the 
experimental results~\cite{rf:itoh,rf:albenque,rf:xu} is explained 
by appropriately taking into account the inhomogeneous structure 
of SC fluctuation.

\begin{figure}[htbp]
  \begin{center}
\includegraphics[width=7.5cm]{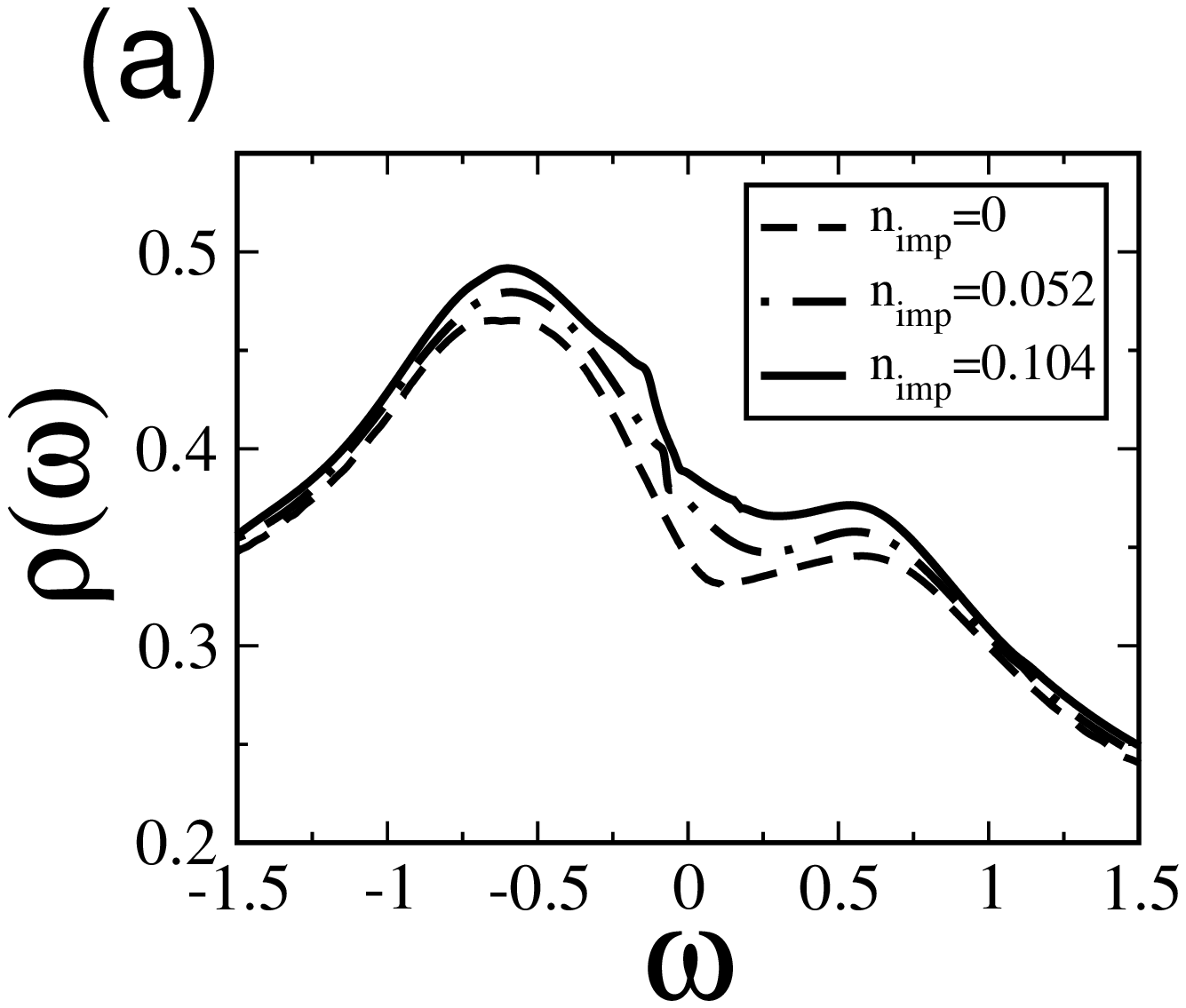}
\hspace{3mm}
\includegraphics[width=7.5cm]{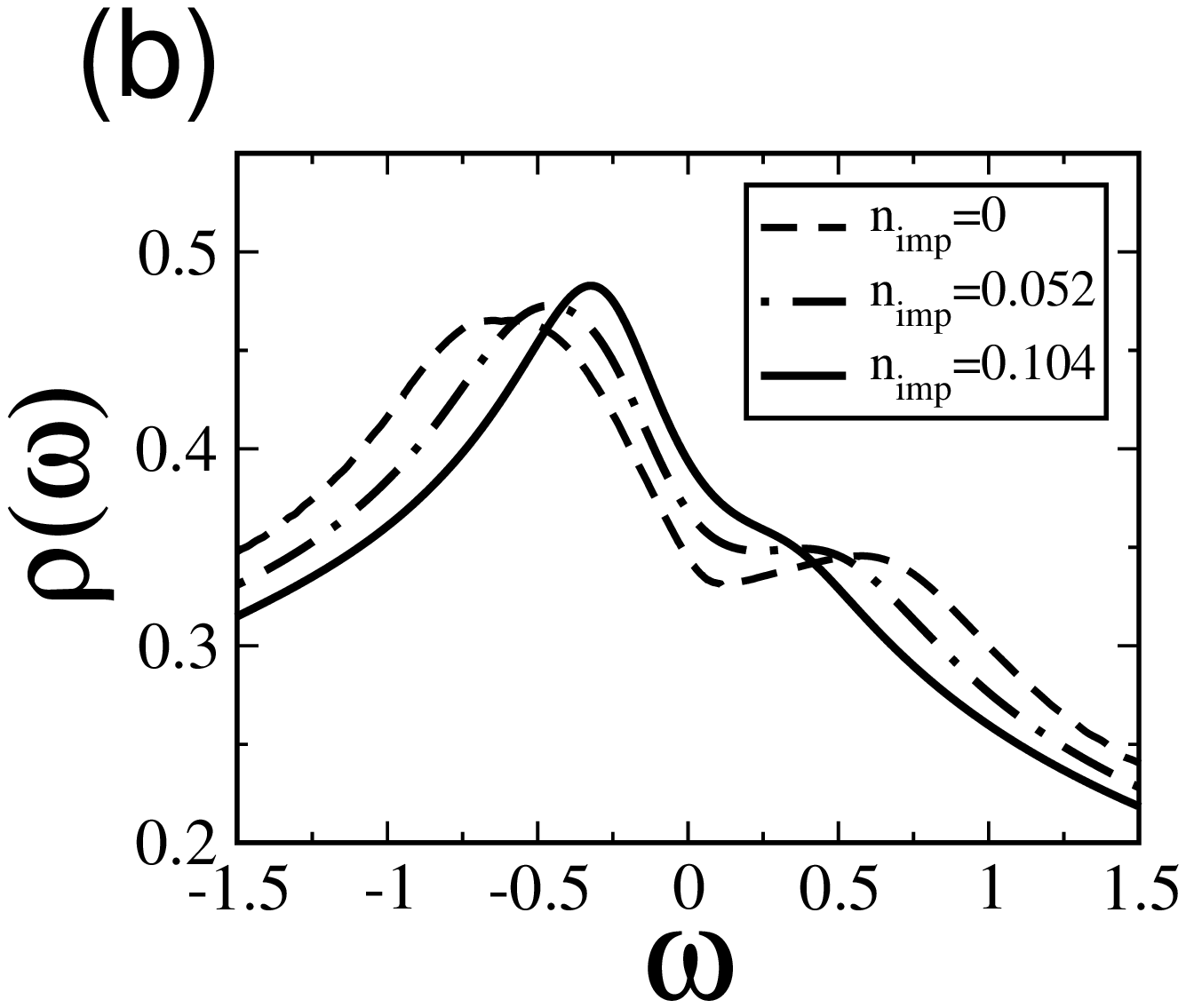}
\caption{The spatially averaged DOS $\rho(\omega)$ for $V=-1.5$. 
(a) The results of the formulation in \S2.1.
We take more than 20 samples to take the random average. 
(b) The results of SCTMA. 
             }
    \label{fig:rspacefldos}
  \end{center}
\end{figure}

 The experiments~\cite{rf:itoh,rf:albenque,rf:xu} have shown that 
the pseudogap is robust for the disorder even though the 
transition temperature is significantly decreased. 
 In order to illuminate this feature, we show the 
single particle DOS at $T=T_{\rm c}$. 
 We calculate the local DOS $\rho(\hat{r},\omega)$ in each sample 
which is defined as, 
\begin{eqnarray}
\rho(\hat{r},\omega) = 
-\frac{1}{\pi} {\rm Im} G^{\rm R}(\hat{r},\hat{r},\omega), 
\label{eq:local-DOS} 
\end{eqnarray} 
where the retarded Green function is obtained by the Pade 
approximation. 
 By taking the spatial average and random average, we obtain 
the DOS per Cu sites as, 
\begin{eqnarray}
\rho(\omega) = 
<\frac{1}{N_{0}}\sum_{r}\rho(\hat{r},\omega)>_{\rm r}.  
\label{eq:DOS} 
\end{eqnarray}
 As shown in Fig.~7(a), the pseudogap in the DOS is not significantly 
destroyed even though the \Tc is remarkably reduced. 
 This result should be contrasted to that of SCTMA where 
the pseudogap is smeared by the same amount of 
disorder (Fig.~7(b)). 
 Thus, the robust pseudogap indicated by the experiments 
is attributed to the role of microscopic inhomogeneity.

\begin{figure}[htbp]
  \begin{center}
\includegraphics[width=7.5cm]{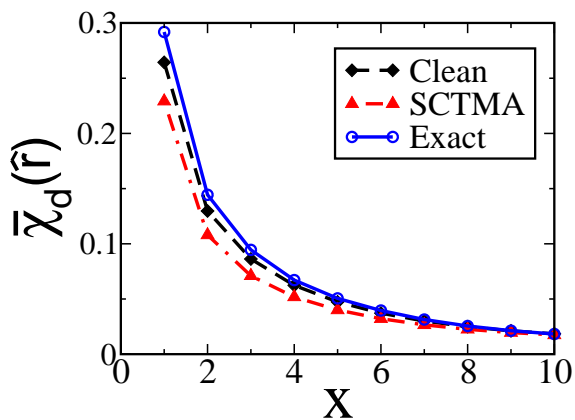}
\caption{
The correlation function of SC order parameter 
$\bar{\chi}_{\rm d}(\hat{r})$. 
 We show the spatial dependence along the diagonal line 
$\hat{r} = (x,x)$. The solid line is the result of the calculation 
formulated in \S2.1 and the dash-dotted line shows the result of SCTMA 
for $n_{\rm imp}=0.104$. The dashed line is the result in the 
clean limit. We choose $V=-1.5$ and the temperature which satisfies 
the criterion eq.~(\ref{eq:Tc-criterion}). 
 Note that the correlation functions at large $x$ are similar to 
each other owing to the criterion. 
             }
    \label{fig:correlation}
  \end{center}
\end{figure}

 In order to understand the results in Figs.~6 and 7 more clearly, 
we again stress that the short range SC correlation develops even 
though the long range coherence is disturbed by the localization of 
order parameter. 
 Because we have defined the \Tc as the temperature where the 
long range correlation begins to develop, the short range correlation 
at that temperature is enhanced by the disorder. 
 This feature is clearly shown in Fig.~8 which shows the correlation 
function of SC order parameter, namely $\bar{\chi}_{\rm d}(\hat{r})$. 
 The pseudogap in the short coherence length superconductor is robust 
for the disorder since the pseudogap is mostly induced by the short range 
correlation.~\cite{rf:yanasereview} 
 We note that the qualitatively different results are obtained in 
the SCTMA where the short range correlation is reduced by the 
disorder. This is because the coherence length increases owing to 
the decrease of \Tc and the long range correlation 
tends to develop. 
 This feature is consistent with Fig.~7(b), but an artifact of the SCTMA.

 It should be noted that the long range correlation is still weak 
in Fig.~8. This means that the actual long range coherence is 
achieved at the temperature lower than \Tc in our definition. 
 In order to discuss the true long range order, we have to take the 
weak three dimensionality into account. 
 The nature of dimensional crossover from 2D to 3D has been investigated 
in the clean systems. Then, the qualitative behaviors of \Tc do not 
depend on the phenomenological criterion.~\cite{rf:yanasePGVC} 
 However, the nature of long range order is not clear in the 
highly disordered system. 
 It is expected from Figs.~3 and 4 that the long range order is 
triggered by the phase coherence between each ``clean region'' 
like granular superconductivity.~\cite{rf:vinokur} 
 The result in Fig.~8 is consistent with this picture although 
the calculation with $N = 31 \times 31$ is not sufficient to describe 
the phase correlation between each granular. 
 We have confirmed that the behaviors in Figs.~6 and 8 become notable 
as increasing the size of calculation. 
 It is an interesting future problem to develop a theory on the crossover 
from the homogeneous superconductivity to the granular superconductivity.

\section{Pseudogap around Single Impurity}

 In \S3, we have investigated the role of microscopic inhomogeneity 
on the macroscopic properties of $d$-wave short coherence length 
superconductor. 
 We briefly discuss the spatial dependence in the atomic scale 
in this section.

 Here, we focus on the spatial dependence of the electronic DOS 
in the presence of strong point disorders like Zn impurities. 
 Because the most significant spatial dependence is the variation 
around the disorder, we show the DOS around the single impurity. 
 The pseudogap around the single impurity has been investigated 
by the NMR measurements.~\cite{rf:itohZn,rf:jullien,rf:alloul,
rf:macfarlane} 
 We discuss the relevance of our results for these experiments.

\begin{figure}[htbp]
  \begin{center}
\includegraphics[width=7.5cm]{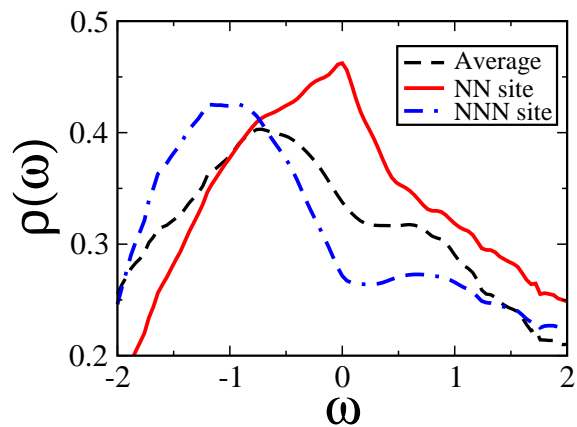}
\caption{
The LDOS around the single impurity. 
We show the LDOS at the nearest neighbor (NN) site (solid line) 
and next nearest neighbor (NNN) site (dot-dashed line) in addition to 
the spatially averaged DOS (dashed line). 
             }
    \label{fig:localdos2d}
  \end{center}
\end{figure}

 In this section, we adopt the self-consistent T-matrix approximation 
for $N = 31 \times 31$ sites.  
 Figure~9 shows the local DOS (LDOS) at the nearest neighbor (NN) 
and next nearest neighbor (NNN) sites where the deviation from 
the averaged DOS is most significant. 
 The temperature is chosen to be slightly above \Tc and therefore 
the pseudogap opens in the averaged DOS as shown by the dashed line. 
 It is clearly shown that the pseudogap is destroyed 
at the NN site. This is because the Cooper pairing 
is forbidden between the impurity site and NN site. 
 We have confirmed that the $d$-wave order parameter is remarkably 
suppressed at the NN site. 
 The destroyed pseudogap at the NN site is consistent with the 
NMR measurements which have shown the Curie-Weiss enhancement 
of $1/T_{1}T$ around the non-magnetic 
impurity.~\cite{rf:itohZn,rf:jullien,rf:alloul,rf:macfarlane} 
 The neutron scattering measurements have also shown the 
destruction of superconducting gap in the magnetic 
excitation.~\cite{rf:sidis,rf:kofu}

\begin{figure}[htbp]
  \begin{center}
\includegraphics[width=7.5cm]{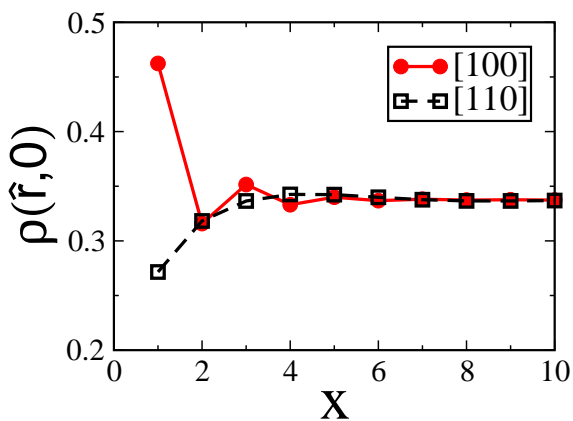}
\caption{
The spatial dependence of the LDOS along the [100] 
and [110] directions from the impurity site. 
We define the impurity site $\hat{r}=(0,0)$ and show 
the LDOS $\rho(\hat{r},0)$ at $\hat{r}=(x,0)$ (circles) and 
$\hat{r}=(x,x)$ (squares), respectively. 
             }
    \label{fig:localdosline}
  \end{center}
\end{figure}

 Figure~9 shows that the pseudogap in the DOS at the NNN site is 
larger than that in the averaged DOS. 
 Thus, the spatial dependence of the DOS is non-monotonic. 
 We show the LDOS $\rho(\hat{r},0)$ at the Fermi energy 
along [100] and [110] directions in Fig.~10. 
 The DOS remarkably oscillates along [100] direction 
while the DOS shows a long wave length oscillation along [110] direction. 
 It should be noted that the SC order parameter is monotonically 
suppressed around the impurity site and shows no oscillation. 
 Thus, the suppression of the DOS along the [110] direction is not 
attributed to the enhancement of the SC correlation. 
 Actually, the oscillation of the DOS is a kind of the quasi-particle 
interference effect~\cite{rf:QIE} which reflects the shape of 
the Fermi surface.

 We have analyzed the magnetic excitations in the disordered 
$t$-$t'$-$U$-$V$ model in order to take into 
account the competing anti-ferromagnetic fluctuation~\cite{rf:kontani} 
in addition to the SC fluctuation. 
 Then, we found that the spatial dependences of static quantities 
are quite different from that of NMR $1/T_{1}T$.  
 The staggard spin susceptibility is monotonically enhanced 
near the impurity. 
 This enhancement of the anti-ferromagnetic correlation induces the 
rapid oscillation of uniform spin susceptibility which has been 
observed in the Knight shift measurements.~\cite{rf:ouazi} 
 This spatial dependence of the static anti-ferromagnetic correlation can 
be qualitatively described by the Ginzburg-Landau theory which assumes 
the competition between the anti-ferromagnetism and superconductivity. 
 In contrast to that, the $1/T_{1}T$ shows a spatial dependence similar 
to the electronic DOS in Fig.~10, which can not be described in the 
Ginzburg-Landau theory. 
 This is consistent with the experimental result on the $1/T_{1}T$ which 
shows the length scale shorter than the Ginzburg-Landau correlation 
length.~\cite{rf:itohanalysis} 
 The details will be shown in the future publication.~\cite{rf:yanasetUV}

 When we distribute many strong point disorders like Figs.~3 and 4, 
the Fourier transformed DOS (FTDOS) is dominated by the deviation 
at the NN and NNN sites. 
 Then, no clear structure appears in the FTDOS around 
$q=(\pi/4,0)$. 
 In contrast to that, the peak appears in the FTDOS around 
$q=(\pi/4,0)$ (not shown) for the weak random disorder. 
 The latter is similar to the experimental observation 
in the pseudogap state,~\cite{rf:vershinin} although the structure 
around $q=(\pi/4,0)$ is smeared in the high frequency region.

 We note that the nanoscale inhomogeneity of DOS 
observed in the STM~\cite{rf:pan,rf:mcelroy,rf:sugimoto} 
is not clearly reproduced in this calculation. 
 We consider that the huge and broad 
gap structure in the weakly superconducting region is due to the 
correlation competing with the $d$-wave 
superconductivity.~\cite{rf:hirschfeldmf} 
 A candidate is the disordered magnetism observed in the neutron,  
$\mu$SR and NMR measurements.~\cite{rf:keimer,rf:wakimoto,rf:bernhard,
rf:sanna,rf:ishidaPS}

\section{Discussion}

\subsection{Summary and discussion}

 We have investigated the disorder-induced microscopic inhomogeneity 
of SC order parameter and its effects on the pseudogap in 
short coherence length superconductors with 
$d$-wave symmetry. 
 We find that the SC order parameter is highly localized 
by a small amount of point disorders and the nanoscale inhomogeneity 
like a granular superconductor appears. 
 Owing to the spatially localized SC correlation, the long range 
coherence hardly develops while the short range correlation is 
enhanced. 
 As a result, the critical region of superconductivity is 
{\it enlarged} by the disorder in contrast to the SCTMA based on the 
Abrikosov-Gorkov theory.~\cite{rf:chen,rf:kudo} 
 Then, the pseudogap induced by the SC fluctuation is robust for the 
disorders.

 One of the typical short coherence length superconductors is 
the high-\Tc superconductor. 
 The under-doped cuprates satisfy three conditions to realize 
the microscopic inhomogeneity, namely the short coherence length, 
non-$s$-wave symmetry and quasi-two dimensionality. 
 Actually, the nanoscale inhomogeneity has been observed in the 
STM measurements~\cite{rf:pan,rf:mcelroy,rf:sugimoto} and also in the 
Josephson plasma resonance measurement.~\cite{rf:shibauchi} 
 We note that the disorder induces the inhomogeneity of SC order 
parameter without remarkable charge inhomogeneity. 
 The results of NMR and transport measurements~\cite{rf:itoh,
rf:albenque,rf:xu,rf:andoAGbreak} indicating the breakdown of 
Abrikosov-Gorkov theory are consistent with the microscopic inhomogeneity 
induced by the disorder. 
 We understand these experimental results in a coherent way 
by appropriately taking the spatial dependence into account.

 It should be stressed that there are intrinsic disorders in 
high-\Tc cuprates. 
 Therefore, our results indicate the nanoscale inhomogeneity 
in the non-disorder-doped sample. 
 Then, it is expected that the superconducting transition in 
the highly under-doped region is triggered by the phase 
coherence of nanoscale granular regions. 
 This picture is consistent with the Uemura plot~\cite{rf:uemura} 
which has implied the importance of phase fluctuation as pointed out 
by Emery and Kievelson.~\cite{rf:emery} 
 The recent experiments have also indicated the two dimensional phase 
fluctuation above \Tcf.~\cite{rf:kitano,rf:ong,rf:corson}

 The nature of the SI transition in the highly under-doped region is 
beyond the scope of this paper. However, it is reasonable 
to consider that the disordered magnetism occurs as shown by the 
experiments.~\cite{rf:keimer,rf:wakimoto,rf:bernhard,rf:sanna} 
 We will discuss this topic in another publication.~\cite{rf:yanasetUV}

\subsection{Discussion on a variety of SI transitions}

 A variety of SI transitions have been observed in 
strongly correlated electron systems. 
 Here, we briefly discuss how the differences can be understood by 
considering the role of disorder. 
 The importance of disorder in the under-doped cuprates is illuminated 
by the following discussion.

 The nature of SI transition is different between the series of 
high-\Tc cuprates. 
 The disordered magnetism has been observed most clearly in the La-based 
systems where the CuO$_2$ layer is significantly affected by the 
disorder. 
 In contrast to that, the recent work on the multi-layer 
cuprates has shown the abrupt change from the superconducting state to the 
anti-ferromagnetic metallic state at finite temperatures.~\cite{rf:mukuda} 
 The latter is consistent with the theoretical results based on the 
FLEX and FLEX+T-matrix approximations in the clean 
system.~\cite{rf:yanasereview} 
 The authors have claimed that the clean CuO$_2$ layer is realized 
in this compound. ~\cite{rf:mukuda} 
 Then, the absence of the disordered magnetism and/or the 
anti-ferromagnetic insulating state implies that the disorder plays 
an essential role for the SI transitions in the other cuprates.

 The organic superconductor $\kappa$-(BEDT-TTF)$_2$X~\cite{rf:kanoda} 
is another candidate for the short coherence length $d$-wave 
superconductor.~\cite{rf:yanasereview,rf:jujo} 
 It is expected that the system is relatively clean because the 
material is free from the dopant disorders. 
 The first order phase transition from the superconductivity to the 
anti-ferromagnetism indicates that the organic superconductor 
$\kappa$-(BEDT-TTF)$_2$X is analogous with the multi-layer high-\Tc 
cuprates.

 The electron-hole asymmetry of the SI transition in cuprate 
superconductors is understood by considering the difference of 
coherence length. The result of the FLEX+T-matrix approximation has shown 
that the coherence length is much longer in the electron-doped region 
than in the hole-doped region.~\cite{rf:yanaseFLEXPG} 
 This is the reason why the pseudogap due to the SC fluctuation is not 
observed in the electron-doped cuprates.~\cite{rf:zhengedope,
rf:yamadakurahashi,rf:almitage} 
 Then, the abrupt change from the superconducting state to the 
anti-ferromagnetic state~\cite{rf:mfujita} is attributed to the 
absence of microscopic inhomogeneity which is unlikely in the long 
coherence length superconductor.

 A variety of SI transitions accompanying the magnetism have been 
discovered in the heavy fermion materials.~\cite{rf:kitaoka} 
 It is reasonable to consider that the multi-orbital effect 
plays an essential role for the phase transition from the 
superconductivity to the magnetism. 
 In some cases, the superconductivity and magnetism 
are induced by the different orbitals. 
 Then, the coupling between the superconductivity and magnetism 
is weak and the co-existent state can be stabilized.

\subsection{Plan of the future study}

 It is a challenging future work to describe the SI transition 
on the basis of the disorder-induced nanoscale 
inhomogeneity.~\cite{rf:trivedi} 
 Some theoretical developments are needed for this aim. 

 First, it is important to take into account the competition 
between the superconductivity and anti-ferromagnetism. 
 This competition induces the spatially inhomogeneous structure of 
superconducting region and magnetic region.~\cite{rf:dagotto} 
 We have examined the spatial structure on the basis of the 
$t$-$t'$-$U$-$V$ model which includes the on-site repulsion. 
 Then, we found that the magnetic correlation is enhanced in the 
``dirty region'' where the superconducting correlation is destroyed 
by the disorder. 
 The results will be shown in the future publication.~\cite{rf:yanasetUV}

 Second, it is necessary to take into account the quantum fluctuation 
to describe the SI transition at zero temperature although we have 
investigated the thermal fluctuation in this paper. 
 One of the promising strategies is to derive the phenomenological 
models such as the Ginzburg-Landau model or the XY model where 
the long range behaviors are illuminated.~\cite{rf:ghosal} 
 We are planning to microscopically derive a phenomenological model 
which is relevant in the spatially coexistent state of the 
superconductivity and anti-ferromagnetism.

\section*{Acknowledgments}

 The authors are grateful to T. Adachi, A. Fujimori, K. Fujida, M. Fujita, 
M. Ido, K. Ishida, Y. Itoh, H. Kitano, Y. Koike, H. Kontani, K. Kudo, 
N. Momono, H. Mukuda, M. Ogata, T. Shibauchi, K. Yamada and T. Yoshida 
for fruitful discussions.  
 Numerical computation in this work was partly carried out 
at the Yukawa Institute Computer Facility. 
 The present work has been partly supported by a Grant-In-Aid for Scientific 
Research from the Ministry of Education, Culture, Sport, Science and 
Technology of Japan.


\begin{thebibliography}{9}
\bibitem{rf:comment}
In order to avoid confusions, we denote the self-consistent t-matrix 
approximation for the disorder as SCTMA in the following part while 
the T-matrix approximation on the SC fluctuation is 
denoted in extenso.
\end{thebibliography}
\end{document}